\documentclass[runningheads]{llncs}

\newcounter{final}
\usepackage{graphicx}
\usepackage[utf8]{inputenc}
\usepackage[T1]{fontenc}
\usepackage{subcaption}
\usepackage{hyperref}
\usepackage{color}

\usepackage[nolist]{acronym}
\usepackage{xcolor}
\usepackage{siunitx}
\usepackage{tikz}
\usetikzlibrary{decorations.pathreplacing,decorations.markings,arrows.meta}
\usepackage{pgfplots}

\newcommand{\ie}{\textit{i.\,e.}}
\newcommand{\AD}[1]{\ifnum \value{final}=0{\color{brown}AD: #1}\fi}
\newcommand{\SD}[1]{\ifnum \value{final}=0{\color{red}SD: #1}\fi}
\newcommand{\MM}[1]{\ifnum \value{final}=0{\color{cyan}MM: #1}\fi}
\newcommand{\GH}[1]{\ifnum \value{final}=0{\color{green}GH: #1}\fi}
\newcommand{\SH}[1]{\ifnum \value{final}=0{\color{orange}SH: #1}\fi}
\newcommand{\SP}[1]{\ifnum \value{final}=0{\color{olive}SP: #1}\fi}
\newcommand{\JX}[1]{\ifnum \value{final}=0{\color{magenta}JX: #1}\fi}
\newcommand{\MB}[1]{\ifnum \value{final}=0{\color{purple}MB: #1}\fi}
\newcommand{\instr}[1]{\ifnum \value{final}=0{\color{blue}Instructions: #1}\fi}

\usepackage{xspace}
\newcommand{\psmpi}{ParaStation~MPI\xspace}

\renewcommand{\url}[1]{}
\newcommand{\doi}[1]{}

\begin{document}

\title{Coupling Complementary Simulations for Combined Performance and Energy Optimization}
\titlerunning{Coupling Complementary Simulations for Combined Perf. and Energy Opt.}

\author{Adel Dabah\inst{1}\orcidID{0000-0001-9175-469X} \and Gregor Häfner\inst{3}\orcidID{0000-0002-1750-493X} \\ \and
Sonja Happ\inst{2}\orcidID{0000-0002-1858-3641} \and
Simon Pickartz\inst{2}\orcidID{0000-0002-6316-6396} \and
Marcus Müller\inst{3}\orcidID{0000-0002-7472-973X} \and
Andreas Herten\inst{1}\orcidID{0000-0002-7150-2505}}
\authorrunning{A. Dabah et al.}
\institute{Jülich Supercomputing Centre, Forschungszentrum Jülich, Germany \email{\{a.dabah,a.herten\}@fz-juelich.de} \and
ParTec AG, Munich, Germany \\
\email{\{sonja.happ,pickartz\}@par-tec.com}
\and
Institute for Theoretical Physics, Georg August University Göttingen, Germany\\ %Friedrich-Hund-Platz 1, Göttingen, 37077, Germany\\
\small \email{gregor.haefner@uni-goettingen.de, mmueller@theorie.physik.uni-goettingen.de }
}

\maketitle

%----------------------------------------------------------------------------------------
% Abstract
%---------------------------------------------------------------------------------------

\begin{abstract}
Polymer simulations are among the most computationally demanding workloads in soft-matter research, often requiring days of execution and high energy consumption to achieve physically meaningful results. In this work, we address these challenges through the coupling and optimization of two complementary simulation frameworks: the Uneyama–Doi Model (UDM) and the SOft coarse-grained Monte Carlo Acceleration (SOMA). UDM efficiently propagates concentration fields at the continuum level, while SOMA resolves chain-scale thermal fluctuations via particle-based Monte Carlo dynamics. Each model was individually optimized for GPU execution using kernel fusion, memory coalescing, asynchronous random-number generation yielding up to \qty{70}{\percent}~(UDM) and \qty{80}{\percent}~(SOMA) performance improvement. The coupling is performed through our proposed coordinator library that orchestrates data exchange and synchronizes time-stepping across multiple GPUs. Further management of coupling workload distribution enabled a 13$\times$ overall speedup and 24.5$\times$ reduction in total energy usage compared to the SOMA baseline, \ie, \qty{96}{\percent} energy saving. The proposed hybrid approach maintains the same scientific fidelity while drastically reducing the computational and energy footprint, showcasing the potential of energy-aware, cross-application co-design for sustainable high-performance simulations.
\end{abstract}


\section{Introduction}
Simulating collective phenomena in polymer systems at realistic scales remains a grand challenge in computational materials science. Despite advances in computational resources, even coarse-grained polymer simulations continue to demand multi-day execution times and substantial energy consumption due to the high dimensionality and stochastic nature of their physical models~\cite{muller2013computational}. Achieving shorter time-to-solution without sacrificing model fidelity is therefore essential for sustainable High-Performance Computing~(HPC) simulations. 

To date, the majority of research on energy efficiency in supercomputing has focused on low-level hardware control, such as Dynamic Voltage and Frequency Scaling (DVFS)~\cite{mei2016dissecting,mei2017survey}, runtime system coordination and resource management strategies, like power capping and multi-objective scheduling~\cite{etinski2012parallel}. While these approaches are important, they primarily optimize energy through infrastructure-level control. In contrast, our strategy focuses on coupling complimentary applications. To our knowledge, this is an under-explored area of research for achieving effective energy efficiency.
%While most existing approaches for energy savings in supercomputing %primarily focus on runtime coordination, hardware optimization and %resource allocation, 

In this work, we target this challenge by coupling two complementary polymer simulation models: \ac{UDM} and \ac{SOMA}. UDM captures macroscopic concentration field dynamics through partial differential equations, enabling efficient large-domain simulations but lacking chain-level statistics~\cite{uneyama_density_2005,hafner2026gpu}. In contrast, SOMA resolves molecular-scale configurations using Monte Carlo~(MC) sampling, providing accuracy at the cost of extremely high computational demand~\cite{schneider2019multi}.

To overcome the limitations of each model while minimizing the energy footprint, both codes were first  optimized independently on A100 and H100 GPUs. For UDM, grouping small kernels, overlapping random-number generation, and fusing compute steps reduced global memory operations by up to \qty{87}{\percent}, achieving a \qty{70}{\percent} runtime improvement. For SOMA, porting initialization and validation to GPU, restructuring memory access, and improving data-reuse for key kernels improved throughput by \qty{80}{\percent}  and reduced the energy consumption by \qty{50}{\percent} compared to the baseline SOMA.  

Beyond these independent optimizations, the two models were coupled using our proposed coordinator library that enables concurrent execution of the two models across multiple GPUs.
It facilitates data communication and synchronization between continuum and particle models. The coupling framework runs UDM as a parent application. Choosing the region, in which the continuum model is expected to show suboptimal accuracy, the parent application starts a child SOMA application to model only this subdomain. Then, the models run concurrently, and, periodically communicating, the high-fidelity results are integrated into the parent application. Profiling using NVIDIA Nsight Systems~\cite{nvidia_nsight_systems} reveals minimal communication overhead and identifies synchronization gaps, which are mitigated through adaptive domain partitioning.

The final coupled multi-scale UDM–SOMA system maintains similar physical fidelity as standalone SOMA simulations while achieving a $13\times$ speedup and $24\times$ reduction in total energy consumption as compared to the baseline SOMA code on two JUWELS Booster nodes with four NVIDIA A100 GPUs. This \qty{96}{\percent} energy reduction illustrates that energy-efficient scientific computing can be achieved not only through hardware optimization but also through intelligent cross-model coupling.

%\SH{The following paragraph needs an update once writing is done.}
The rest of the paper is organized as follows: \hyperref[sec:opt]{Section~\ref*{sec:opt}} gives an overview of the two complementary simulation models and optimizations performed. \hyperref[sec:coupling]{Section~\ref*{sec:coupling}} introduces the proposed coordinator library and general view of the coupling mechanism. 
\hyperref[sec:results]{Section~\ref*{sec:results}} summarizes the obtained results and discussions. \hyperref[sec:lessons]{Section~\ref*{sec:lessons}} provides the lessons learned and finally, \hyperref[sec:conclusion]{Section~\ref*{sec:conclusion}} concludes this paper. 

\section{Overview and Optimizations}
\label{sec:opt}
This section gives an overview of the two complementary simulation models, \ac{UDM} and \ac{SOMA}, and introduces the \acp{GPU} optimizations performed for ensuring efficient hardware utilization.  

\ac{UDM} is a continuum model that propagates local concentration fields over time by numerically integrating coupled partial differential equations, which captures the diffusive dynamics of each monomer species, driven by solvent-nonsolvent exchange and equilibration within a free energy landscape~\cite{uneyama_density_2005,hafner2026gpu}. This approach offers significant advantages in computational efficiency and scalability, allowing for the simulation of large domains with fewer resources. However, this efficiency comes at the cost of reduced accuracy, as the model lacks explicit polymer-chain configuration statistics and neglecting the thermal noise.

\ac{SOMA} offers a particle-based framework to study multi-component polymer systems at a coarse-grained level, designed for high-performance computers~\cite{Blagojevic2023Dec,schneider2019multi}. It implements the Single-Chain-in-Mean-Field (SCMF) algorithm using a soft, coarse-grained polymer model to investigate phase separation and self-assembly in soft matter systems. As a result, it requires substantial computational resources, which can be efficiently parallelized using the quasi-instantaneous-field approximation for non-bonded interactions. The time evolution of the system is modeled through bond-force-biased \ac{MC} propagation of polymer beads within the \ac{SOMA} program, allowing the model to naturally capture polymer dynamics with thermal fluctuations, an essential aspect of the underlying physical processes. 

Coupling of these two approaches as a multi-scale solution, using the coordinator library (see~\autoref{sec:coordinator}), combines the strengths of both methods:  the computational efficiency of  \ac{UDM} and the physical accuracy of the particle-based \ac{SOMA}. Before coupling and to fully utilize modern GPUs, each model is optimized independently for \ac{GPU} execution to improve its performance and reduce memory bottlenecks and energy consumption.  After coupling, additional optimization ensures balanced workload distribution and reduced synchronization overhead across multiple GPUs. This end-to-end optimization, before and after coupling, increases simulation throughput and keeps large-scale hybrid simulations both efficient and sustainable. 

\subsection{UDM Optimization}
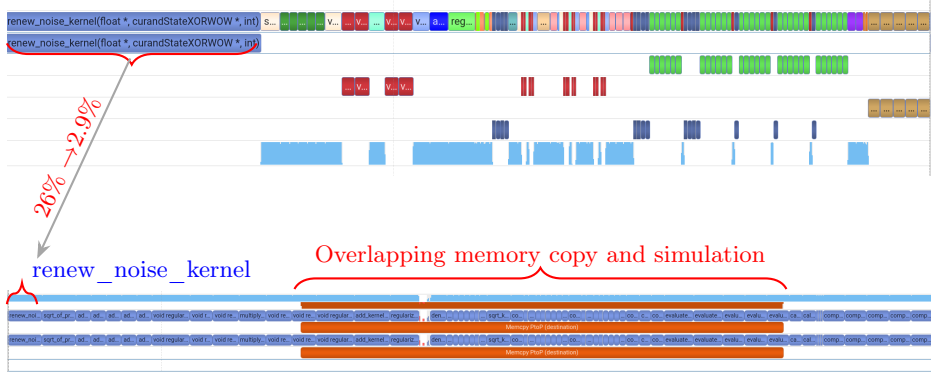
\begin{figure}[tb]
    \centering
    \begin{tikzpicture}%[scale=1, transform shape]
        % --- Top image ---
        \node[anchor=south west, inner sep=0] (top) at (0,2.60)
            {\includegraphics[width=\linewidth]{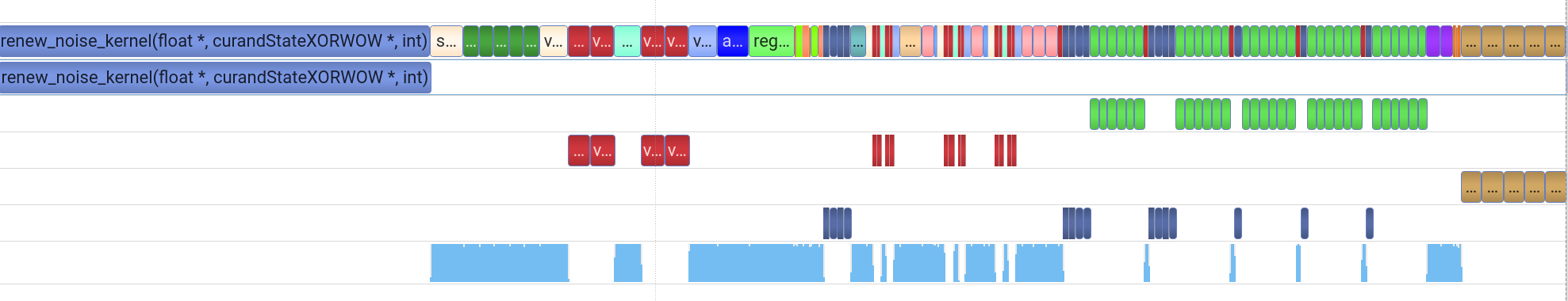}};
        
        % Bracket for renew_noise kernel (adjust coordinates as needed)
        \draw[decorate,decoration={brace,amplitude=8pt},thick,red]
            (3.3,4.35) -- (0.,4.35)
            node[midway,above=8pt,blue]{};
        
        % --- Bottom image ---
        \node[anchor=south west, inner sep=0] (bottom) at (0,0)
            {\includegraphics[width=\linewidth]{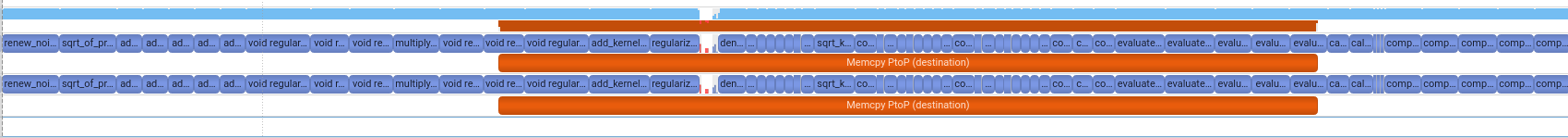}};
        
        % Bracket for first kernel in bottom image
        \draw[decorate,decoration={brace,amplitude=8pt},thick,red]
            (0.0,0.9) -- (00.4,0.9)
            node[midway,below=-20pt,xshift=45pt,blue]{renew\_noise\_kernel };
             
            node[midway,sloped,above=3pt,red]{ };

             \draw[->,thick,gray!70,>=Stealth]
            (1.63,4.15) -- (0.4,1.5)
            node[midway,sloped,above=3pt, yshift=-4pt, xshift=-2pt, red]{26\% \textrightarrow 2.9\%};
            
            \draw[decorate,decoration={brace,amplitude=8pt},thick,red]
            (3.8,1.1) -- (10.3,1.1)
            node[midway,below=-20pt,red]{Overlapping memory copy and simulation};
    \end{tikzpicture}

    \caption{UDM kernels in one time step using Nsight Systems profiling tool (top) and overlapping \ac{RNG} with the UDM simulation (bottom).}
    \label{fig:udm_rng}
\end{figure}

To optimize \ac{UDM} for \ac{GPU} execution, we first conduct a detailed analysis of its computational workflow.  These insights guide targeted optimizations to improve parallel efficiency, memory utilization, and overall simulation throughput. For this purpose, the top part of \autoref{fig:udm_rng} shows the NVIDIA Nsight Systems profile for a single \ac{UDM} time step.

%\begin{figure}[htbp!]
    %\centering
    %\includegraphics[width=\linewidth]{figures/UDM_kernels.png}
    %\caption{UDM kernels in one time step using Nsight Systems profiling tool.}
    %\label{fig:udm}
%\end{figure}

In the top part of \autoref{fig:udm_rng}, less than \qty{1}{\percent} of data transfers occur between the \ac{CPU} and \ac{GPU}. Each time step executes up to 112 short kernels in less than  \qty{10}{\milli\second}, dominated by the \texttt{renew\_noise\_kernel} function~(\qty{27}{\percent}) and 24 Fourier transforms (\qty{16}{\percent}). To reduce kernel launch overhead and memory traffic, multiple small kernels are grouped into larger ones. This kernel fusion yields a \qty{20}{\percent} performance gain, reducing the kernel count from 112 to 69, and lowering global memory operations by up to \qty{33}{\percent} for reads and \qty{87}{\percent} for writes, enabling the memory subsystem to reach \qty{90}{\percent} throughput.

The \texttt{renew\_noise\_kernel} is responsible for \ac{RNG} in \ac{UDM} and is further optimized by extracting \ac{RNG} to a dedicated GPU and overlapping the generation with  simulation. This reduces the share of the function of the runtime from \qty{26}{\percent} to \qty{2.9}{\percent} (bottom part of \autoref{fig:udm_rng}) and yields an additional \qty{30}{\percent} speedup, resulting in a total \qty{70}{\percent} improvement over the baseline \ac{UDM} application without optimizations.  %\SH{The baseline should be defined somewhere.}
The approach generates random numbers for $L$ upcoming steps in advance and transfers them asynchronously, ensuring they are ready when needed. 

\begin{figure}[tb]
\centering
%\begin{minipage}{0.5\textwidth}
\begin{tikzpicture}
\begin{axis}[
    height=4cm, width=7.5cm,
    ybar,
    bar width=15pt, % Slightly reduced bar width for clarity
    ylabel={Performance (TPS)},
    ylabel near ticks,
    % Define the Symbolic X-Coordinates and use them precisely
    symbolic x coords={A100 Baseline, A100 opt., H100 opt.},
    %x tick label style={
        %rotate=0, % Rotate labels for better fit
        %anchor=east, % Anchor the label for cleaner rotation
        %font=\footnotesize % Use small font size
    %},
    nodes near coords,
    nodes near coords align={vertical},
    ymin=0,
    ymax=145, % Increased ymax to fit the annotation nodes clearly
    enlarge x limits=0.3,
    ymajorgrids,
    xtick distance=1.0,  
    grid style={dashed,gray!40},
    %tick label style={font=\tiny},
    %label style={font=\tiny},
    %title={UDM Performance Comparison Across GPUs},
    % Remove the legend style if no legend is explicitly added (see notes below)
]

% Define the coordinates based on the Symbolic X-Coords list
\addplot[fill=gray!50] coordinates {(A100 Baseline, 38) (A100 opt., 65) (H100 opt., 96)};
%\addplot[fill=cyan!60] coordinates {(A100 opt., 65)};
%\addplot[fill=orange!80] coordinates {(H100 opt., 96)};
%\addplot[fill=green!70!black] coordinates {(GH200 opt., 120)};

% --- Performance Increase Annotations (Calculated based on Baseline: 38) ---

% UDM_opt (A100): (64-38)/38 = 68.4%
\node[font=\bfseries, anchor=south, align=center]     at (axis cs:A100 opt., 65) [yshift=10pt] {+70\%};

% UDM_opt (H100): (75-38)/38 = 97.4%
% Corrected syntax: {Symbolic Name, Y-coordinate}
\node[font=\bfseries, anchor=south, align=center]     at (axis cs:H100 opt., 95) [yshift=10pt] {+155\%};

% UDM_opt (GH200): (120-38)/38 = 215.8%
% Corrected syntax: {Symbolic Name, Y-coordinate}
%\node[font=\bfseries, anchor=south, align=center]
%    at (axis cs:GH200 optimised, 120) [yshift=10pt] {\tiny{+216\%}};
\end{axis}
\end{tikzpicture}
%\end{minipage}
%\hfill
%\begin{minipage}{0.4\textwidth}
 
 \caption{UDM code performance in \ac{TPS} for the baseline and optimized versions, tested across various NVIDIA GPU architectures. The percentages indicate the speedup relative to the A100 baseline.}
 \label{fig:udm_gpu_comparison}
 %\end{minipage}

\end{figure}
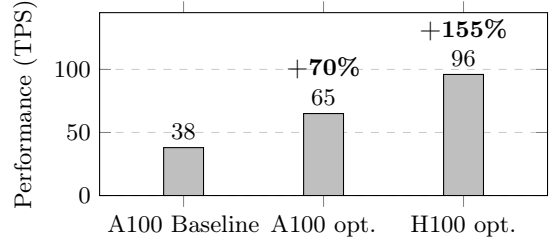
    
Moreover, \autoref{fig:udm_gpu_comparison} shows steady \ac{UDM} performance gains across NVIDIA GPUs, with the optimized H100 achieving a \qty{155}{\percent} speedup over the A100 baseline through combined software and hardware improvements.

\subsection{SOMA Optimization}
Following the same approach as for \ac{UDM}, the \ac{SOMA} application~\cite{herten2024application} is profiled to identify major performance bottlenecks. The initialization and testing phases originally consume about \qty{235}{\second}, dominated by generating bead coordinates and validation functions. Offloading these routines to the GPU reduces their runtime to under \qty{2}{\second}, eliminating repeated overhead when coupling with the continuum model.
\begin{figure}[tb]
    \centering
    \includegraphics[width=0.99\linewidth]{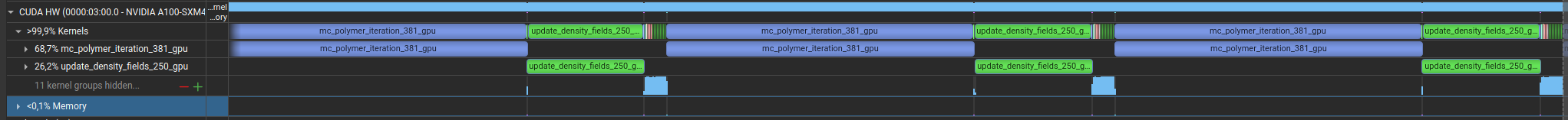}
    \caption{Profile of the SOMA application with two major kernels consuming 95\% of execution time}
    \label{fig:soma_kernels}
\end{figure}

As depicted in \autoref{fig:soma_kernels}, two main kernels dominated the execution time:  \texttt{mc\_polymer\_iteration} (\qty{65}{\percent}), which performs biased trial moves and energy evaluations, and \texttt{update\_density\_fields} (\qty{25}{\percent}), which updates particle density fields. Both kernels are optimized by improving memory coalescing, caching frequently used variables, and adjusting launch configurations to reduce thread divergence and register pressure. As a result, memory throughput nearly doubles, and kernel runtimes are improved by \qty{40}{\percent}. %\SH{Also for the SOMA optimizations the baseline should be defined somewhere.}

%\begin{minipage}{0.49\textwidth}
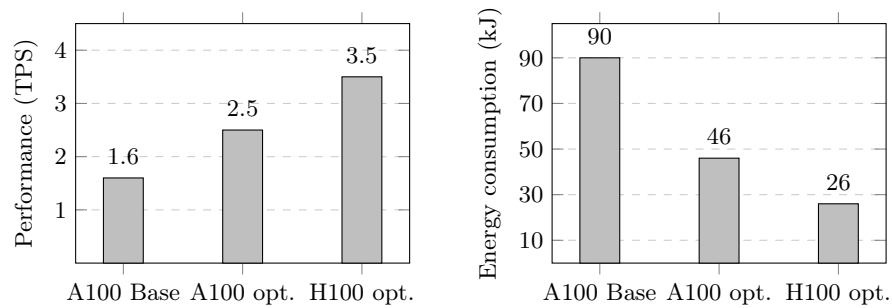
\begin{figure*}[tb]
\begin{subfigure}[t]{0.49\textwidth}
\centering
\begin{tikzpicture}%[scale=0.6]
\begin{axis}[
        ybar,
    bar width=15pt,
    enlarge x limits=0.2,
    ymajorgrids,
    ymin=0, ymax=4.5,
    xtick={2,3,4},
    xticklabels={{A100 Base},{A100 opt.},{H100 opt.}},
    %xticklabel style={%xshift=0.7cm,
        %rotate=0, % Rotate labels for better fit
        %anchor=east,
        % Anchor the label for cleaner rotation
        %font=\footnotesize % Use small font size
        %},
    ytick={1,2,3,4,5},
    ylabel={Performance (TPS)},
    ylabel near ticks,
    height=4.75cm, width=6cm,
    nodes near coords,
    nodes near coords align={vertical},
    point meta=y,
       grid style={dashed,gray!40},
    nodes near coords style={anchor=south, yshift=2pt}
]
\addplot[fill=gray!50] coordinates {(2,1.6) (3,2.5) (4,3.5)};
%\addplot[fill=cyan!60] coordinates {(3,2.5)};
%\addplot[fill=orange!80] coordinates {(4,3.5)};
%\addplot[fill=blue!70] coordinates {(5,5.3)};
\end{axis}

% Example annotation lines (adjusted for positioning)
%\draw[<->,blue,thick] 
%  (6.3,1) -- (6.3,3.8)
%  node[midway, right, blue]{\small $3\times$};
\end{tikzpicture}
\end{subfigure}
~
%\end{figure}
%\end{minipage}
\hfill
\begin{subfigure}[t]{0.49\textwidth}

%\begin{minipage}{0.49\textwidth}
    \centering
 % --- POWER CONSUMPTION BARPLOT ---
\begin{tikzpicture}%[scale=0.6]
\pgfkeys{/pgf/number format/.cd,1000 sep={\,}}
\begin{axis}[
     ybar,
    bar width=15pt,
      ymajorgrids,
     enlarge x limits=0.2,
    ymin=0, ymax=105,
    xtick={2,3,4},
    xticklabels={{A100 Base},{A100 opt.},{H100 opt.}},
    %xticklabel style={xshift=0.7cm,  rotate=40, % Rotate labels for better fit
        %anchor=east, % Anchor the label for cleaner rotation
        %font=\footnotesize},
    ytick={10,30,50,70,90},
    ylabel={Energy consumption (kJ)},
    ylabel near ticks,
    height=4.75cm, width=6cm,
    nodes near coords,
    nodes near coords align={vertical},
    point meta=y,
       grid style={dashed,gray!40},
    nodes near coords style={anchor=south, yshift=2pt}
]
\addplot[fill=gray!50] coordinates {(2,90) (3,46) (4,26)};
%\addplot[fill=cyan!60] coordinates {(3,46)};
%\addplot[fill=orange!80] coordinates {(4,26)};
%\addplot[fill=blue!70] coordinates {(5,18.8)};

\end{axis}
% Example annotation lines
%\draw[<->,blue,thick] 
%  (7.1,1.2) -- (7.1,3.6)
 % node[midway, right, blue]{\small $3.8\times$};

\end{tikzpicture}
%\end{minipage}
\end{subfigure}
 \caption{Performance in \ac{TPS} and energy consumption in kJ of the \ac{SOMA} application across GPU generations. }

\label{fig:energy2}

\end{figure*}

%\SH{Terminology: we should use power consumption, energy consumption and energy efficiency in a more consistent manner.}

\autoref{fig:energy2} shows the performance and energy efficiency of \ac{SOMA} using one \ac{GPU}. From the A100 baseline (SOMA without optimizations) to the H100 \acp{GPU}, the \ac{SOMA} application achieved over a $2\times$ speedup and a more than three times reduction in energy consumption, highlighting the benefits of combined software tuning and architectural advances. Note that the impact of optimizations in a multi-node setup is reduced by communication overhead at each step of the simulation.
\begin{figure}[!h]
\centering
\begin{subfigure}{0.48\textwidth}
  \centering
  \begin{tikzpicture}
    \begin{axis}[
        width=\textwidth, height=4.5cm,
        xlabel={GPUs}, ylabel={Speedup (vs 4 GPUs)},
        xtick={4,8,12,16}, ymin=0.5, ymax=4.5,
        legend pos=north west,
        legend style={font=\scriptsize,draw=none,fill=none},
        grid=both, grid style={dashed,gray!25},
    ]
    \addplot[mark=*,blue,thick] coordinates {(4,1) (8,1.6) (12,1.74) (16,2.33)};
    \addlegendentry{Actual}
    \addplot[mark=square,gray,dashed] coordinates {(4,1) (8,2) (12,3) (16,4)};
    \addlegendentry{Ideal scaling}
    \end{axis}
  \end{tikzpicture}
  \caption{Speedup – higher is better.}
\end{subfigure}\hfill
\begin{subfigure}{0.48\textwidth}
  \centering
  \begin{tikzpicture}
    \begin{axis}[
        width=\textwidth, height=4.5cm,
        xlabel={GPUs}, ylabel={Steps per kJ (norm.)},
        xtick={4,8,12,16}, ymin=0.5, ymax=1.5,
        legend pos=north east,
        legend style={font=\scriptsize,draw=none,fill=none},
        grid=both, grid style={dashed,gray!25},
    ]
    \addplot[mark=*,teal,thick] coordinates {(4,1.0) (8,0.8) (12,0.58) (16,0.58)};
    \addlegendentry{Actual}
    \addplot[mark=square,gray,dashed] coordinates {(4,1.0) (8,1.0) (12,1.0) (16,1.0)};
    \addlegendentry{Ideal (constant efficiency)}
    \end{axis}
  \end{tikzpicture}
  \caption{Steps per kJ – higher is better.}
\end{subfigure}
\caption{SOMA scaling on A100 GPUs: from 4 to 16 GPUs, speedup reaches only $2.33\times$ (parallel efficiency $\approx 58\%$). Energy per step increases by $72\%$, and energy efficiency (steps/kJ) drops by $42\%$. Adding GPUs makes each simulation step \emph{less} energy efficient, motivating our coupled approach that uses only 4 GPUs.}
\label{fig:soma_scaling_three_panels}
\end{figure}
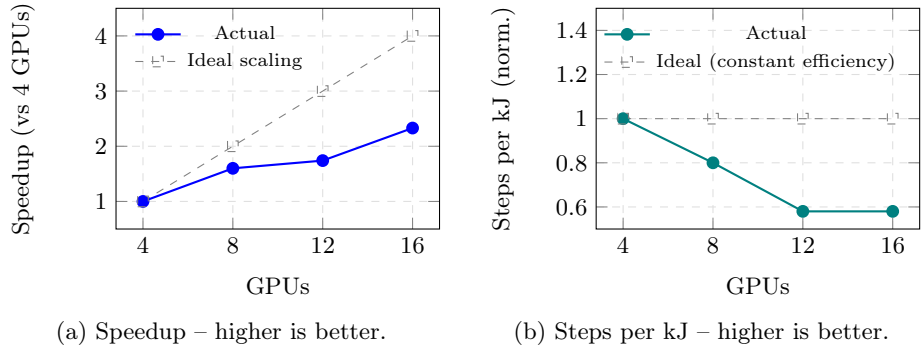

\autoref{fig:soma_scaling_three_panels} shows that scaling standalone SOMA from 4 to 16 GPUs  leads to significant parallel efficiency reduction due to communication and synchronization overheads. More critically, it \emph{increases} energy per timestep, as idle GPUs still draw power. This exposes a fundamental limit of scaling the expensive particle model alone. 
Our coupled UDM–SOMA approach bypasses this limit by replacing most of the expensive work with the efficient continuum UDM on just 4 GPUs, yielding higher speed and far lower energy per step.
    
%    \begin{minipage}{0.53\textwidth}
%\autoref{fig:tps_gpu_comparison} presents the \acf{TPS} performance of the SOMA application when scaling the number of nodes (4 GPUs per node). The performance increases from \qty{2}{TPS} on 4 GPUs to \qty{4.66}{TPS} on 16 GPUs, corresponding to only about a 2.2$\times$ speedup and a parallel efficiency of only \qty{58.25}{\percent}, \ie, using four times more computational resources and higher energy consumption for a relatively modest performance gain. Indeed, density-field communication overhead between all \acp{GPU}  remains the dominant scalability bottleneck. Adding more hardware resources is not a sustainable path. Instead, improving energy efficiency requires algorithmic and software-level innovations as demonstrated by the multi-scale approach combining \ac{UDM} and \ac{SOMA}.
%\end{minipage}

\section{Coupling Scheme} 
\label{sec:coupling}

% \item Performance trade-offs in communication / synchronization 
% \end{itemize}

The coupling scheme performs a concurrent multi-fidelity modeling based on the \ac{UDM} and \ac{SOMA} applications using the coordinator library. 

\subsection{Coordinator Library}
\label{sec:coordinator}
\begin{figure}[bt]
  \centering
  \includegraphics[width=\textwidth]{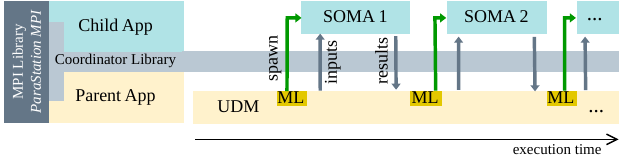}
  \caption{Overview of coordinator library integration into MPI library (\psmpi) and parent (UDM) and child (SOMA) applications. The workflow at execution time including spawn of SOMA via ML-based decision making and data exchange between parent and child applications is shown on the right.}
  \label{fig:coordinator-lib}
\end{figure}

Both, \ac{SOMA} and \ac{UDM} are \ac{MPI} applications.
A comparison of their coupling requirements with the technical possibilities and state-of-the-art in MPI simulation coupling suggested to implement their coupling
as a lightweight MPI-based \emph{coordinator} library.
As shown in \autoref{fig:coordinator-lib}, the computationally less expensive continuum model \ac{UDM} acts as the \emph{parent} application for the coupling, while the computationally more expensive \ac{SOMA} model is spawned as \emph{child} application.

For spawning the child application, the coordinator library leverages the \ac{MPI} spawning interface \verb|MPI_COMM_SPAWN| and the capabilities of integrating new \ac{MPI} processes into \ac{MPI} communication via an inter-communicator.
This ensures that the spawn mechanism is portable to other HPC applications and not limited to the simulation coupling use case presented in this work.
In order to decide when to spawn a child application and what data to send to the child application, the coordinator library calls the \ac{ML}-based decision-making callback ~\cite{busch2025machine}, which is a multilayer perceptron (MLP) model called from the parent application.  It takes descriptors of the current and previous concentration fields from UDM and outputs the expected discrepancy between continuum and particle results. The subdomain boundaries are then identified via thresholding and post-processed to enforce the zero-flux constraint. Training data were generated from paired UDM–SOMA simulations. Inference overhead is negligible; details are given in~\cite{busch2025machine}.
The outputs of this callback provide the spawn decision as well as the input data for the child application to be spawned.

The input data is sent to the child application using \ac{MPI} communication procedures and an inter-communicator.
Similarly, once the child application has finished its computation, results are sent back to the parent application for further processing and integration into the overall coupled simulation progress.
Through this exchange of data using \ac{MPI} communication parent and child applications are synchronized to the required extent.

The parent application may spawn several instances of the child application consecutively based on the need and the outcome of the decision-making callback (see for example the spawn of~ ``SOMA 2'' in \autoref{fig:coordinator-lib}).
Thereby, the resources used by the child application can be tailored to the use case dependent algorithmic and computational requirements that emerge at execution time. 
This fine-granular control over resource usage helps minimizing the energy footprint of the overall coupled simulation since the child application may only be spawned at times when it is required for the input data that it really needs to process.

For the simulation experiments presented in this work, \psmpi~\cite{Suarez:909256} was used together with the coordinator library which was compiled and linked against \psmpi.
To provide additional spawning parameters in the \verb|MPI_COMM_SPAWN| procedure, such as the number and type of nodes or \acp{GPU} to be used by the child application, non-\ac{MPI} standard features were added to \psmpi.
The parallel execution environment can then take these additional parameters into account to spawn the child application processes on the desired resources.
Using the coordinator library with other \ac{MPI} libraries works in general, however the additional non-standard spawn configuration features may not be available.

\subsection{Coupling}

Given a morphology (polymer structure), \ie, the concentration fields ${\phi_\alpha(\mathbf{r},t)}$, a coupling decision and subsequent spawn or communication step are performed periodically every $t_\mathrm{sync}$. At the first synchronization time, the \ac{UDM} communicates the current and previous morphologies to the \ac{ML} model~\cite{busch2025machine}, which predicts the subdomain boundaries in the $z$-direction for the particle-based \ac{SOMA} simulation. To satisfy the zero polymer flux constraint, boundaries are extended if required. The \ac{UDM} then spawns a \ac{SOMA} simulation via the coordinator library (see \autoref{fig:coordinator-lib}) and receives an \ac{MPI} inter-communicator for data exchange. A virtual HDF5 image containing simulation parameters, subdomain concentration fields, and boundary fluxes is transmitted to initialize the \ac{SOMA} run.

Each \ac{SOMA} simulation begins by generating particle positions that reproduce the supplied concentration fields. Molecules are placed within half a mean end-to-end distance of their target positions and equilibrated using an umbrella potential for about half a synchronization period. The boundary zones in the simulation act as transition layers where particles are converted into continuum units (and vice versa) to maintain a constant flow and stable density at the edges of the box.
%The subsequent dynamic simulation enforces fixed boundary fluxes by wall confinement and molecular-type conversions in boundary zones. 
These boundaries (fluxes) and conversion rates are periodically updated through synchronization with the \ac{UDM}. The parent \ac{UDM} concurrently evolves the full domain and replaces the subdomain concentrations with the data received from \ac{SOMA}. If polymer fluxes exceed a threshold, the \ac{SOMA} run is terminated and a new subdomain is spawned. 
Readers may refer to \cite{hafner2026concurrently} for more details about the coupling theory.

\section{Results}
\label{sec:results}

In this section, we present example results of large simulation domains and discuss considerations required for ideal synchronization and the resulting speedup compared to the baseline \ac{SOMA} on 8 A100 GPUs.
All comparisons in this section use NVIDIA A100 GPUs (JUWELS Booster
nodes, $4$ A100s per node, interconnected via NVLink within nodes and 4
HDR-200 InfiniBand across nodes).

We deliberately run the coupled simulation on fewer GPUs (4) than the standalone SOMA baseline (8), making the comparison conservative. This asymmetry is by design: coupling is a way to achieve more with less.

The energy consumption is obtained from the LLview job accounting system~\cite{frings2025supporting}, which reports total node energy (CPU, GPU, memory, and network) at job granularity. The reported energy per time step is calculated as the total job energy divided by the number of simulation steps. 

\begin{figure}[h!]
\centering
\includegraphics[width=0.85\textwidth]{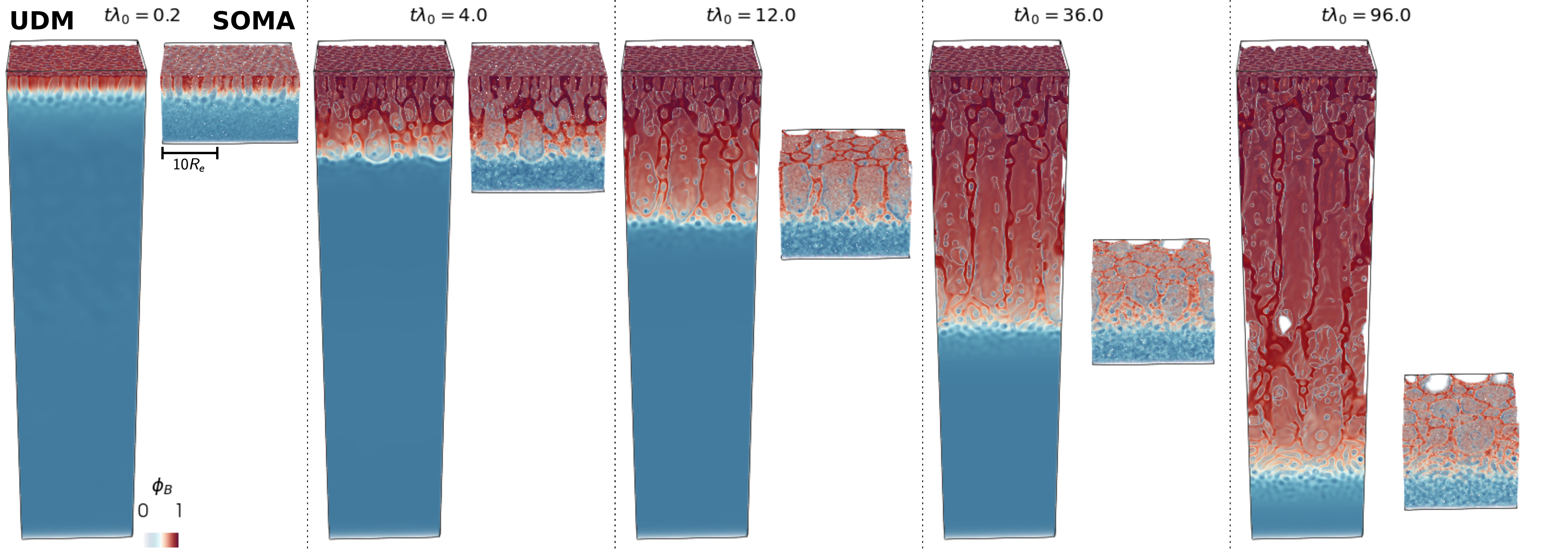}
\caption{Time evolution of the $B$-block concentration $\phi_B(\mathbf{r},t)$ in \ac{UDM} (left) and \ac{SOMA} (right).}
\label{fig:nips-coupled}
\end{figure}

%Using this multiscale scheme, we simulated the \ac{NIPS} process in a large domain $V=L_x\times L_y\times L_z=19.4R_e\times22.4R_e\times78.6R_e$ with $N_x\times N_y\times N_z=194\times224\times786$ grid cells for a total time $T=60\lambda_0^{-1}$. 
Using this multiscale scheme, we simulate the \ac{NIPS} process within a three-dimensional domain of size $V = L_x \times L_y \times L_z = 19.4R_e \times 22.4R_e \times 96R_e$, discretized into $N_x \times N_y \times N_z = 194 \times 224 \times 960$ grid cells. Here, $R_e$ denotes the polymer end-to-end distance, used as a natural unit of length. The $x$ and $y$ plane represents the lateral dimensions of the membrane, while the extended $z$-direction ($L_z$) corresponds to the film-thickness or phase-separation axis. The system is evolved for a total simulation time of $T = 96\lambda_0^{-1}$. 
The synchronization period is $t_\mathrm{sync}=0.16\lambda_0^{-1}$, and each spawned \ac{SOMA} simulation runs for 15 such periods. The simulation was executed on a single JUWELS Booster node, assigning one GPU to \ac{UDM} and three to \ac{SOMA}, resulting in \SI{\sim 20}{\%} \ac{UDM} idle time.

As shown in \autoref{fig:nips-coupled}, the subdomain follows the structure formation front where nonsolvent first phase-separates from the polymer. The  used error prediction guides subdomain placement by tracking the region of highest error, effectively capturing the structure formation front and maintaining the polymer flux below the prescribed threshold. 

\begin{figure}[tb!]
\centering
\includegraphics[width=0.6\linewidth]{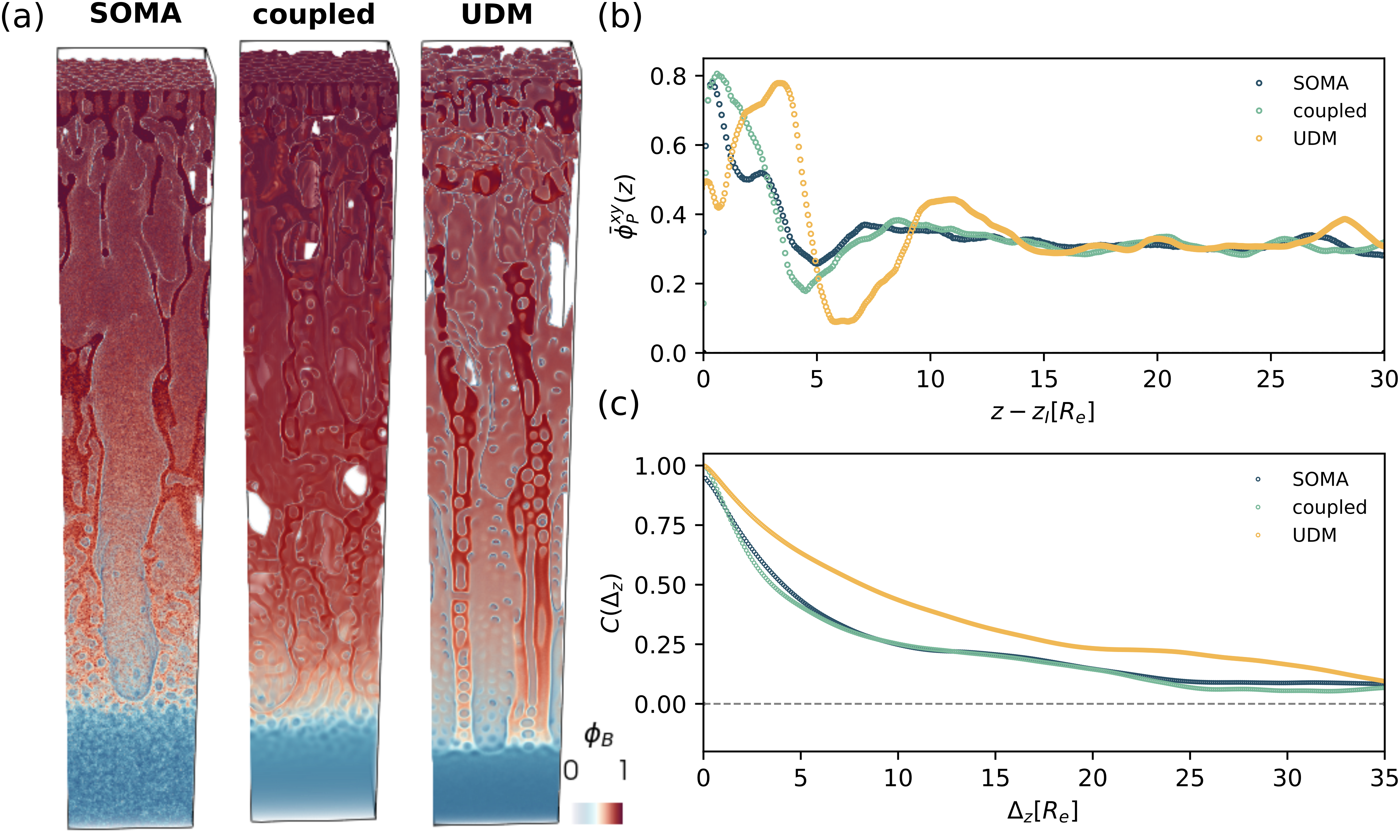}
\caption{Comparison of final $B$-block concentration fields of standalone \ac{SOMA}, coupled and standalone-\ac{UDM} simulation, showing (a) the 3D morphology, (b) the laterally-averaged polymer concentration and (c) vertical correlation.}
\label{fig:comparison_without_cutoff}
\end{figure}

The final output structure represents the vitrified membrane structure, that can be compared to standalone simulations of \ac{SOMA} and \ac{UDM}.
\hyperref[fig:comparison_without_cutoff]{Figure~\ref*{fig:comparison_without_cutoff}} compares the coupled multi-scale simulation to an independent full \ac{SOMA} run ($V=14R_e\times16R_e\times200R_e$, constrained to the smaller $z$-dimension in the plot), as well as to an independent \ac{UDM} simulation ($V=14R_e\times16R_e\times 96R_e$, at increased grid resolution $L_x/N_x=L_y/N_y=L_z/N_z=R_e/15$). This is deliberate to give the
standalone UDM its best achievable accuracy, to see the impact of the coupling. Contrarily to the UDM simulation, the coupled simulation closely reproduces the high-fidelity \ac{SOMA} results with only minor deviations near the interface ($z-z_I\approx5$–$10R_e$) due to residual continuum dynamics. Quantitative analyses of laterally-averaged polymer density $\bar{\phi}_P^{xy}(z)$ and vertical correlation $C(\Delta_z)$ confirm highly-improved agreement of the coupled simulation compared to the standalone \ac{UDM} simulation.
In addition, while \ac{UDM} over-estimates the persistence of the macro-pores, the coupled version successfully resolved this issue bringing the simulation results in line with the SOMA reference.
Note that the three simulations in \autoref{fig:comparison_without_cutoff} use different domain extents in $z$ because each method has different stability requirements: SOMA requires $L_z = 200\,R_e$, whereas the coupled simulation reaches the same converged morphology at $L_z = 96\,R_e$ because the continuum part of the domain absorbs the far-field dynamics. Therefore, we compare each
method at the domain size it requires to be physically valid, not at a common $L_z$. 

%The former quantity indicates increased dissolution of the polymer structure, while the latter indicates an over-estimated persistence of the macro-pores in the continuum model, which is recovered in the coupled simulation. 
%This demonstrates that the coupled approach captures both lateral and vertical structure formation with high fidelity.

\begin{figure}[tb]
    \centering
   
    \begin{tikzpicture}[scale=0.95, transform shape]
        % --- Bottom image ---
        \node[anchor=south west, inner sep=0] (bottom) at (0,0)
            {\includegraphics[width=\linewidth]{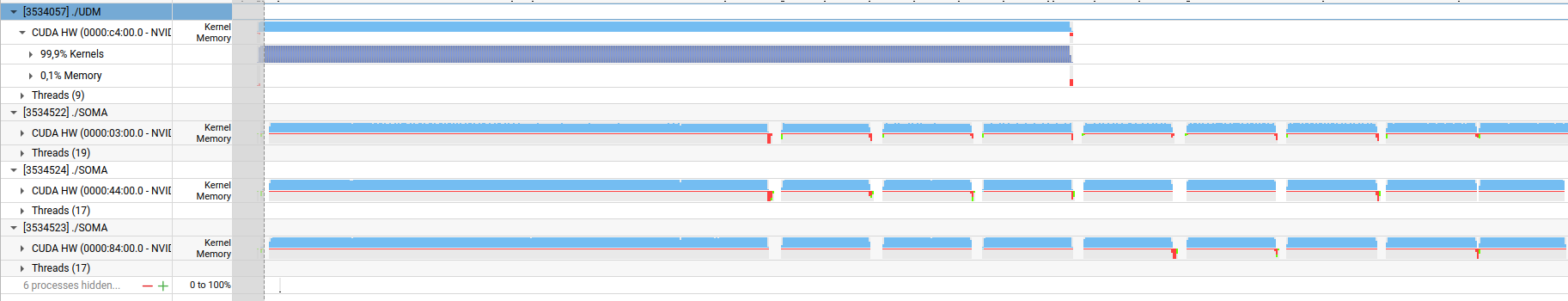}};
        
        % Bracket for first kernel in bottom image
        \draw[decorate,decoration={brace,amplitude=8pt},thick,blue]
            (0.1,0) -- (0.1,1.5)
            node[near start,above=28pt, left=13pt,rotate=90,blue]{\small SOMA  };

            \draw[decorate,decoration={brace,amplitude=8pt},thick,red]
            (8.5,2.2) -- (12,2.2)
            node[midway,below=0pt,red]{\small UDM is waiting for SOMA  };
    \end{tikzpicture}
    \caption{Nsight Systems  profiling of the coupled multi-scale simulation. }
    \label{fig:coupled-delay}
\end{figure}

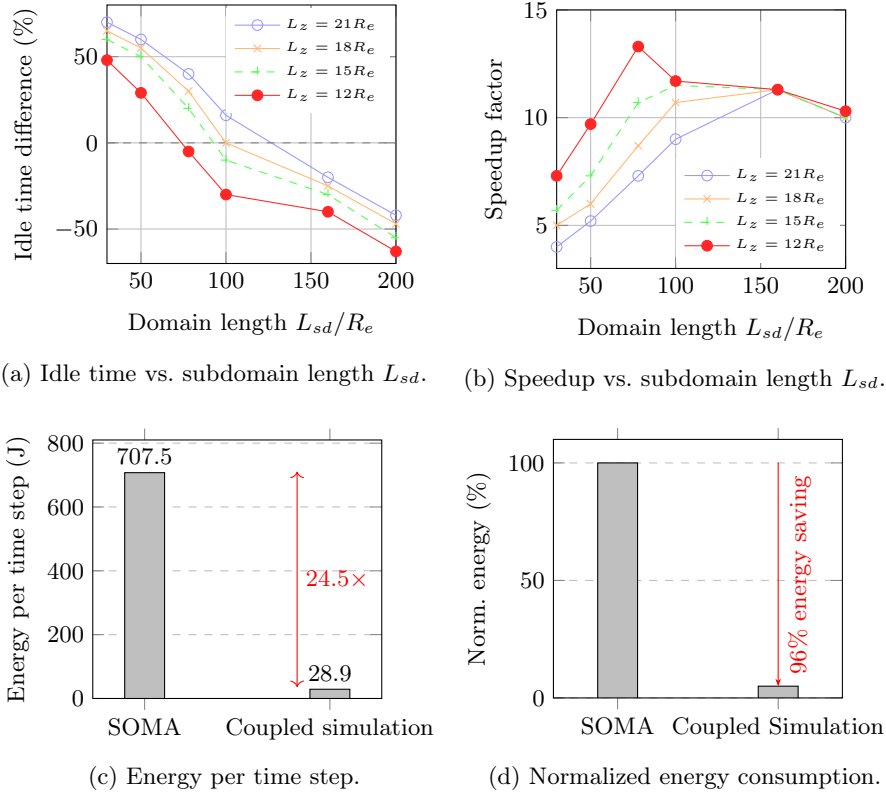
\begin{figure*}[h!]
% --- Left: Idle time and speedup (stacked subfigures) ---
%\begin{minipage}[c]{0.49\linewidth}   % <-- changed from [t] to [c]
\begin{subfigure}[c]{0.49\linewidth}
\centering
\begin{tikzpicture}
\begin{axis}[
width=5.4cm, %\linewidth,
height=5cm,
xlabel={Domain length $L_{sd}/R_e$},
ylabel={Idle time difference (\%)},
ylabel near ticks,
grid=major,
xmin=30, xmax=200,
ymin=-70, ymax=80,
legend style={font=\tiny,draw=none,at={(0.98,0.99)},anchor=north east},
]

\addplot[blue!40,mark=o] coordinates {
(30,70) (50,60) (78,40) (100,16) (160,-20) (200,-42)
};
\addlegendentry{$L_z=21R_e$}

\addplot[orange!55,mark=x] coordinates {
(30,65) (50,55) (78,30) (100,0) (160,-25) (200,-47)
};
\addlegendentry{$L_z=18R_e$}

\addplot[green!70,mark=+,dashed] coordinates {
(30,60) (50,50) (78,20) (100,-10) (160,-30) (200,-55)
};
\addlegendentry{$L_z=15R_e$}

\addplot[red!85,mark=*] coordinates {
(30,48) (50,29) (78,-5) (100,-30) (160,-40) (200,-63)
};
\addlegendentry{$L_z=12R_e$}

\addplot[gray,dashed] coordinates {(30,0) (200,0)};

\end{axis}
\end{tikzpicture}
\caption{Idle time vs.\ subdomain length $L_{sd}$.}
\label{fig:idle_time_subdom_length}
\end{subfigure}
%\end{minipage}
\vspace{0.4cm}
%\begin{minipage}{0.49\linewidth} 
\begin{subfigure}[c]{0.49\linewidth}
\centering
\begin{tikzpicture}
\begin{axis}[
width=5.4cm, %\linewidth,
height=5cm,
xlabel={Domain length $L_{sd}/R_e$},
ylabel={Speedup factor},
ylabel near ticks,
grid=major,
xmin=30, xmax=200,
ymin=3, ymax=15,
legend style={font=\tiny,draw=none,at={(0.98,0.02)},anchor=south east},
]

\addplot[blue!40,mark=o] coordinates {
(30,4.0) (50,5.2) (78,7.3) (100,9.0) (160,11.3) (200,10.0)
};
\addlegendentry{$L_z=21R_e$}

\addplot[orange!55,mark=x] coordinates {
(30,5.0) (50,6.0) (78,8.7) (100,10.7) (160,11.3) (200,10.0)
};
\addlegendentry{$L_z=18R_e$}

\addplot[green!70,mark=+,dashed] coordinates {
(30,5.7) (50,7.3) (78,10.7) (100,11.5) (160,11.3) (200,10.0)
};
\addlegendentry{$L_z=15R_e$}

\addplot[red!85,mark=*] coordinates {
(30,7.3) (50,9.7) (78,13.3) (100,11.7) (160,11.3) (200,10.3)
};
\addlegendentry{$L_z=12R_e$}

\end{axis}
\end{tikzpicture}
\caption{Speedup vs.\ subdomain length $L_{sd}$.}
\label{fig:speedup_subdom_length}
\end{subfigure}

%\end{minipage}%
\hfill
% --- Right: Energy consumption bar chart ---
%\begin{minipage}[c]{0.49\linewidth}   % <-- changed from [t] to [c]
\begin{subfigure}[c]{0.49\linewidth}
\begin{tikzpicture}

\begin{axis}[
    ybar,
    bar width=15pt,
    ymin=0,
    ymax=810,
    ylabel={Energy per time step (J)},
    ylabel near ticks,
    symbolic x coords={SOMA, Coupled simulation},
    xtick=data,
    nodes near coords,
    width=5.4cm,
    height=5cm,   ymajorgrids=true,
      grid style=dashed,
    enlarge x limits=0.28,
]

\addplot[fill=gray!50] coordinates {
    (SOMA,707.5)
    (Coupled simulation,28.9)
};

\end{axis}

% Arrow annotation
\draw[<->,red]
(2.7,3) -- (2.7,0.15)
node[midway,right] {\small $24.5\times$};

\end{tikzpicture}
\caption{Energy per time step.}

%\caption{Average energy per time step for the two applications. The optimized implementation reduces the energy per time step by approximately $24.5\times$.}
\label{fig:energy_per_step}
\end{subfigure}
%\end{minipage}
%\hspace{0.4cm}
%\newline
%\begin{minipage}{0.49\linewidth} 
\begin{subfigure}[c]{0.49\linewidth} 
\begin{tikzpicture}
  \begin{axis}[
      width=5.4cm,
     height=5cm,
      ybar,
      bar width=15pt,
      ymin=0,
      ymax=110,
      ylabel={Norm. energy (\%)},
      xtick={1,2},                         % numeric positions
      xticklabels={SOMA, Coupled Simulation}, % labels for those positions
      %nodes near coords,
      %every node near coord/.append style={font=\footnotesize, /pgf/number format/fixed, anchor=south},
      enlarge x limits=0.4,
      ylabel near ticks,
      ymajorgrids=true,
      grid style=dashed,
      tick label style={font=\small,rotate=0},
      label style={font=\small}
       ]
    % Data bars: x=1 for Baseline, x=2 for Coupled
    \addplot[fill=gray!50] coordinates {(1,100) (2,5)};
    %\addplot[fill=blue!70!cyan] coordinates {(2,5)};

    \draw[red,  -{Stealth[length=1mm]}, line cap=round]
      (axis cs:2,100) --
      node[midway, right=8pt, text=red, font=\small, rotate=90, anchor=center]{96\% energy saving}
      (axis cs:2,5);
  \end{axis}
\end{tikzpicture}
\caption{Normalized energy consumption.}
\label{fig:norm_energy_per_timestep}
\end{subfigure}
%\end{minipage}

\caption{Coupled multi-scale simulation idle time (\ref{fig:idle_time_subdom_length}) and overall speedup (\ref{fig:speedup_subdom_length}) as functions of subdomain length $L_\mathrm{sd}$, demonstrating up to 13$\times$ speedup and 96
\% energy saving (\ref{fig:energy_per_step} and \ref{fig:norm_energy_per_timestep}) compared to the baseline \ac{SOMA} on two nodes, \ie, 8 A100 \acp{GPU}.}
\label{fig:combined_energy_speedup}
\end{figure*}

\autoref{fig:coupled-delay} shows an excerpt of the NVIDIA Nsight Systems timeline for the coupled multi-scale simulation, illustrating the concurrent execution of the conti\-nuum-based \ac{UDM} and the particle-based \ac{SOMA} components. The figure also highlights the minimal impact of communication overhead. However, it reveals synchronization delay, causing to waste computational resources. The idle periods observed occur when the continuum UDM simulation finishes its time steps faster than the corresponding particle-based SOMA simulation. By varying the total domain size \(L_z\) and subdomain length \(L_{\mathrm{sd}}\), we can control this computational imbalance as shown in \autoref{fig:combined_energy_speedup}.% smaller subdomains reduce SOMA's workload and minimize  UDM's wait time. %However, as we show next, this comes at the cost of more frequent spawn operations and potentially lower overall speedup.

 \autoref{fig:combined_energy_speedup} shows the impact of varying the simulation domain length $L_\mathrm{z}$  \ie, the thickness dimension of the simulated polymer film  and SOMA subdomain size \(L_{\mathrm{sd}}\) on idle time, and overall performance in our multi-scale coupled simulation.  \autoref{fig:idle_time_subdom_length} plots the UDM idle time percentage for different subdomain sizes~\(L_{\mathrm{sd}}\).  Increasing the subdomain length increases computational cost of a SOMA simulation which is proportional to the number of simulated polymer beads. This impacts the idle time of the UDM process. Two configurations lead to near zero percent idle UDM time at \{\(L_{\mathrm{sd}} = 12R_e, L_{\mathrm{z}} = 78R_e\)\} and \{\(L_{\mathrm{sd}} = 18R_e,  L_{\mathrm{z}} = 100R_e \)\}.  
 
 \autoref{fig:speedup_subdom_length} shows the speedup compared to a full SOMA simulation on two JUWELS Booster nodes with four A100 GPUs each. Indeed, the two configurations from \autoref{fig:idle_time_subdom_length} correspond to the best attained speedups compared to baseline, showing respectively 13$\times$ and 11.5$\times$. The best trade-off occurs at \(L_{\mathrm{sd}} = 18R_e, L_{\mathrm{z}} = 78R_e \), where the coupled simulation achieves both a close to zero UDM idle time, and standard thickness of the simulated polymer film reflecting on the speedup to reach \(13\times\). Increasing the thickness of film beyond the standard value reduces the overall speedup due to UDM taking more time to finish than SOMA. 

\autoref{fig:energy_per_step} and \autoref{fig:norm_energy_per_timestep} present the energy (J) consumption per time step for the baseline SOMA and multi-scale coupled simulation. \autoref{fig:energy_per_step} shows a more than~24$\times$ improvement in energy consumption per time step resulting in up to \qty{96}{\percent} saved energy compared to the full SOMA baseline on eight A100 GPUs. This dramatic saving stems from two factors: (i) the inherent efficiency of the continuum UDM for the simulation domain, and (ii) the reduced idle time when the subdomain is optimally sized, demonstrating that performance and energy efficiency are aligned in this multi-scale approach.

\section{Lessons Learned}
\label{sec:lessons}
We summarize four lessons that we believe transfer beyond this work.
\paragraph{Lesson 1: Adding more computers saves time, not energy.}
A natural reaction to a slow simulation is to run it on more
processors. This shortens the execution time but it does not, in
general, reduce the total energy consumed. The reason is that
once communication between processors starts to dominate, each
processor spends time waiting rather than computing, while still drawing power. 
Formally, if $\eta(N)$ measures the parallel efficiency for $N$ processors 
(ideal is $\eta=1$), then the runtime scales as
$T(N) = T(1)/\left(\eta(N) N\right)$ but the total energy scales as $E(N) \approx \frac{P_{\mathrm{GPU}}\,T(1)}{\eta(N)}$, where $P_{\mathrm{GPU}}$ represents the power consumption per GPU.
Hence, it only grows worse as efficiency $\eta$ drops, which is the case for large numbers of scientific applications. 
In our case, doubling the number of GPUs from $8$ to $16$ made the simulation $1.45\times$ faster but cost $1.72\times$ more energy per simulated time step.
Therefore, energy savings cannot come from speedup alone or efficient hardware, they must also have to come from less workload if possible. This is exactly what our coupled model achieves, by replacing the expensive particle-based calculation with a much cheaper continuum calculation. 

%\paragraph{Lesson 2: The ``cheap-global plus expensive-local'' pattern is generalizable.} 
%Our specific recipe pairs a fast but approximate model that covers the whole simulation with a slow but accurate model that runs only on a small region where accuracy matters. 
%This pattern is not unique to polymer physics. The same structure appears in weather and climate modeling (a coarse global model coupled to a fine-grained
%model over a single storm), in chemistry (classical molecular dynamics coupled to a quantum-mechanical calculation around a reacting bond), and in engineering fluid dynamics (a fast turbulence approximation coupled to a detailed simulation near a critical component). The lightweight MPI-based coordinator library is engineered to support this general pattern.

\paragraph{Lesson 2: Simulations should weigh accuracy against
energy efficiency.}
At present, our framework decides to invoke the expensive model
whenever the cheap model's prediction looks physically unreliable.
This is a sensible criterion but it ignores cost. A more refined
criterion would compare the expected gain in accuracy against the
energy required to obtain it, invoking the expensive model only
when the gain is worth the energy cost. 
Building this cost-awareness into scientific simulations is, in our view, an underexplored direction with a good energy-saving potential.

\paragraph{Lesson 3: ``Faster'' and ``energy-efficient'' are not the same thing,
and should be reported separately.}
Performance in high-performance computing are generally expressed as speedups. 
Energy are usually expressed as percentage reductions. These are related but not equivalent, and reporting only one can be misleading: a code can become faster while becoming \emph{less} energy-efficient, by running more processors
that mostly wait on each other. Three performance indicators are sufficient to
remove the ambiguity: (1) How much faster the code runs (speedup), (2) how
much energy each unit of useful work consumes, and (3) how many units
of work are done per Watt. 
For our scaling experiment, the three performance indicators told visibly different
stories, which is why we report all three. We suggest this triple,
together with full-node power measurements, for example as displayed through integrated tools such as LLview~\cite{frings2025supporting} reports at FZJ-JSC.

\paragraph{Lesson 4: Heterogeneous workloads require heterogeneous workflows} Effective tooling is required in order to combine separate applications into a super-application to solve multi-scale/multi-fidelity scientific challenges collaboratively.
Such super-application use cases are not limited to the polymer physics use case studied in this paper.
They are also conceivable for example in weather and climate modeling (a coarse global model coupled to a fine-grained model over a single storm), in chemistry (classical molecular dynamics coupled to a quantum-mechanical calculation around a reacting bond), and in engineering fluid dynamics (a fast turbulence approximation coupled to a detailed simulation near a critical component).
A lightweight coordinating entity can effectively decouple direct application-application interaction and act as a generalizable translator. Well-defined APIs are essential. Also heterogeneous hardware platforms are enabled through this coordinator, an important feature for upcoming HPC system designs.

 %\paragraph{Lesson 5: Generalization}
 %While other super-applications from Lesson 4 are conceivable, it is an open question if similar performance gains can be achieved generally.
 %The qualitative trend, that algorithmic restructuring has more impact than hardware-only optimization, is the part that likely generalizes. %Thus, it is better to prioritize algorithmic innovation before hardware upgrades.

\section{Conclusion}
\label{sec:conclusion}
This work demonstrates a pathway toward sustainable, energy-aware polymer simulation by combining two algorithmically complementary methods with GPU-specific optimization. By coupling the continuum UDM with the particle-based SOMA, and optimizing both independently and jointly, we achieved a 13$\times$ speedup resulting in 24.5$\times$ less energy consumed per time step and \qty{96}{\percent} reduction in energy consumption while maintaining the same scientific fidelity. The results highlight the high impact of algorithmic design and complementarity compared to GPU-specific optimization alone.   
The coordinator library provides a mechanism for managing hybrid, multi-model workflows on heterogeneous HPC systems. 
The results underline the importance of software co-design and workload orchestration for sustainable simulations, where performance per Watt/Joule has become a key efficiency metric.

Future work will extend the multi-scale idea to other workloads to minimize the energy footprint for large-scale scientific simulations.

\begin{credits}
\subsubsection{\ackname}

This work was funded by the German Federal Ministry of Research, Technology and Space (MExMeMo, grants 16ME0661,16ME0660,16ME0658K). Computing time was provided by the Gauss Centre for Supercomputing (GCS) via the John von Neumann Institute for Computing (NIC) on the JUWELS Booster at JSC. Access to JUPITER and JEDI was granted through the JUPITER Research and Early Access Program. JUPITER is funded by EuroHPC, the German Federal Ministry, and the Ministry of Culture and Science of North Rhine-Westphalia.
%This work was supported by the German Federal Ministry of Research, Technology and Space within project MExMeMo (16ME0661,16ME0660,16ME0658K). The authors gratefully acknowledge the Gauss Centre for Supercomputing~(GCS) for funding this research project by providing computing time through the John von Neumann Institute for Computing (NIC) on the GCS Supercomputers JUWELS Booster at JSC. Access to the JUPITER supercomputer and the JEDI development system was granted through the JUPITER Research and Early Access Program. JUPITER is funded by the EuroHPC Joint Undertaking, the German Federal Ministry of Research, Technology and Space, and the Ministry of Culture and Science of the German state of North Rhine-Westphalia.
\end{credits}


\bibliographystyle{splncs04}
\bibliography{ref}

\begin{acronym}
  \acro{API}{Application Programming Interface}
  \acro{CPU}{Central Processing Unit}
  \acro{DFG}{Deutsche Forschungsgemeinschaft}
  \acro{DFT}{Density-Functional Theory}
   \acro{FFT}{Fast Fourier Transformation}
  \acro{GPU}{Graphics Processing Unit}
  \acro{HPC}{High Performance Computing}
  \acro{JSC}{J\"ulich Supercomputing Centre}
  \acro{MC}{Monte-Carlo}
  \acro{MCS}{Monte-Carlo Step}
  \acro{ML}{Machine Learning}
  \acro{MPI}{Message-Passing Interface}
  \acro{NIPS}{Nonsolvent-Induced Phase Separation}
  \acro{PMI}{Process Management Interface version 1}
  \acro{PMIx}{Process Management Interface for Exascale}
  \acro{RNG}{Random Numbers Generation}
  \acro{SCMF}{Single-Chain-in-Mean-Field}
  \acro{SMC}{Smart Monte-Carlo}
  \acro{SOMA}{SOft coarse grained Monte-carlo Acceleration}
  \acro{TPS}{Time steps Per Second}
  \acro{UDM}{Uneyama-Doi Model}
  \acro{YAML}{Yet Another Markup Language}
\end{acronym}

\end{document}